\RequirePackage{fix-cm}
\documentclass[natbib,smallextended]{svjour3}       
\usepackage{amsmath}
\smartqed  
\usepackage[dvipdfmx]{graphicx}
 \journalname{my journal}
\newcommand{\aap}{{Astron. Astrophys.}}
\newcommand{\apj}{{Astrophys. J.}}

\newcommand{\la}{\mathrel{\hbox{\rlap{\lower.55ex \hbox {$\sim$}}
 \kern-.3em \raise.4ex \hbox{$<$}}}}
\newcommand{\ga}{\mathrel{\hbox{\rlap{\lower.55ex \hbox {$\sim$}}
 \kern-.3em \raise.4ex \hbox{$>$}}}}
\def\gtrsim{\stackrel{>}{{}_\sim}}
\begin{document}

\title{Gamma-ray bursts and Population III stars
}


\author{Kenji Toma \and Sung-Chul Yoon \and Volker Bromm}


\institute{K. Toma \at
  Frontier Research Institute for Interdisciplinary Sciences, Tohoku University, Sendai 980-8578, Japan; Astronomical Institute, Tohoku University, Sendai 980-8578, Japan \\
              \email{toma@astr.tohoku.ac.jp}           
           \and
           S.-C. Yoon \at
              Department of Physics and Astronomy, Seoul National University, Gwanak-ro 1, Gwanak-gu, Seoul, 88026, South Korea
              \email{yoon@astro.snu.ac.kr}           
           \and
           V. Bromm \at
           Department of Astronomy, University of Texas, 2511 Speedway, Austin, TX 78712, USA
           \email{vbromm@astro.as.utexas.edu}
}

\date{Received: date / Accepted: date}

\maketitle

\begin{abstract}
Gamma-ray bursts (GRBs) are ideal probes of the epoch of the first stars and galaxies. We review the recent theoretical understanding of the formation and evolution of the first (so-called Population~III) stars, in light of their viability of providing GRB progenitors. We proceed to discuss possible unique observational signatures of such bursts, based on the current formation scenario of long GRBs. These include signatures related to the prompt emission mechanism, as well as to the afterglow radiation, where the surrounding intergalactic medium might imprint a telltale absorption spectrum. We emphasize important remaining uncertainties in our emerging theoretical framework.
\end{abstract}

\section{Introduction}
\label{intro}

With recent progress in observational cosmology, one of the key frontiers is to understand how the first stars
and galaxies ended the cosmic dark ages a few hundred million years after the Big Bang \citep{Bromm09,Loeb10}. Prior to their emergence, the universe exhibited a simple state, devoid of complex structure, of any elements heavier than lithium, and of high-energy radiation fields. Within $\Lambda$CDM cosmology, the first stars, the so-called Population III (Pop III), are predicted to form at $z \sim 20 - 30$ in dark matter minihalos of mass $\sim 10^6 M_{\odot}$. The formation of the first bona-fide galaxies, being able to host long-lived stellar systems, may be delayed until more massive dark matter halos virialize \citep{Bromm11}. Once the first sources of light have appeared on the cosmic scene, the universe was rapidly transformed through the input of ionizing radiation \citep{Barkana07} and heavy chemical elements \citep{Karlsson13}. The character of this fundamental transition, as well as the assembly process of the first galaxies, crucially depended on the feedback exerted by Pop~III stars \citep{Ciardi05}. The feedback in turn is determined by the initial mass function (IMF) of the first stars \citep{Bromm04,Glover05}. Although important uncertainties remain, the key prediction is that the Pop~III IMF is biased towards high mass, implying a top-heavy distribution \citep{Bromm13}. At least a fraction of the first stars could therefore have collapsed into massive black holes (BHs) at the end of their short lives, and thus provide viable gamma-ray burst (GRB) progenitors.

Upcoming facilities such as the James Webb Space Telescope (JWST), and the next generation of extremely-large telescopes on the ground (GMT, TMT, E-ELT) promise to open up a direct window into the first billion years of cosmic evolution \citep{Gardner06}. Despite their exquisite sensitivity at near-IR wavelengths, even these observatories may not be able to directly probe the first stars, unless they formed in massive clusters \citep{Pawlik11}, or were gravitationally lensed \citep{Rydberg13}. The only opportunity to probe individual Pop~III stars may be to catch them at the moment of their explosive death. This could involve extremely energetic supernova (SN) events, such as hypernovae or pair-instability SNe \citep{Hummel12,Pan12,Whalen13}, or GRBs. The latter fate depends on whether Pop~III stars could give rise to suitable collapsar progenitors, involving rapidly rotating massive stars \citep{MacFadyen99}. Since Pop III stars are predicted to fulfill both requirements (see the discussion below), GRBs are expected to be prevalent at very high redshifts. Indeed, GRBs may play a key role in elucidating primordial star formation, as well as the properties of the early intergalactic medium (IGM), given their extreme intrinsic brightness, both of the prompt $\gamma$-ray emission, as well as that of the prolonged afterglow.

A number of features render GRBs ideal probes of the epoch of first light \citep{Loeb10}: {\it (i)} Traditional sources to observe the high-$z$ universe, such as quasars and Lyman-$\alpha$ emitting galaxies, severely suffer from the effects of cosmological dimming. It was suggested, on the other hand, that GRB afterglows, if observed at a fixed time after the trigger, may exhibit nearly-flat infrared fluxes out to very high $z$ \citep{Ciardi00}. The argument behind this counter-intuitive effect was as follows: A fixed time interval in the observer frame translates into an increasingly early time in the source frame. Such earlier times in turn would sample the rapidly decaying GRB lightcurve at the moment of maximal brightness, thus compensating for the cosmological dimming (increasing luminosity distance). With the realization that such simple power-law decay may not be established until quite late after the trigger, this distance independence of the high-$z$ burst flux now appears too optimistic. {\it (ii)} In the hierarchical setting of cosmic structure formation, earlier times are dominated by lower-mass host systems. The massive hosts required for quasars and bright galaxies therefore are “dying out” at the highest redshifts \citep{Mortlock11}. GRBs, on the other hand, mark the death of individual stars, which can form even in very low-mass systems. {\it (iii)} Finally, Pop~III GRBs would provide very clean background sources to probe the early IGM. Again reflecting the low masses of their hosts, any proximity effect should be much reduced, as ionized bubbles are confined to the immediate vicinity of the Pop~III system; the IGM would thus largely remain unperturbed. In addition, since GRB afterglow spectra can be described as featureless, broken power-laws \citep{Vreeswijk04}, any signature imprinted by absorption and emission events along a given line of sight can be easily discerned. 

The outlook for GRB cosmology, therefore, is bright, provided that we can continue to fly wide-field $\gamma$-ray trigger instruments in space, beyond the end of the {\it Swift} satellite. We already have tantalizing examples of high-redshift bursts with the spectroscopically confirmed GRB~090423 at $z \simeq 8.2$ \citep{Salvaterra09,Tanvir09}, and the photometrically detected GRB~090429B at $z \simeq 9.4$ \citep{cucchiara11}. Future missions, such as SVOM \citep{paul11}, promise to fully unleash the potential of GRBs to probe the early universe. This review is organized as follows. In Section~2, we discuss the recent consensus on the formation of the first stars, with a particular focus on assessing their suitability as GRB progenitors. In Section~3, we continue by summarizing the key stellar evolution physics of these suggested Pop~III progenitors. In Section~4, we discuss promising observational avenues to probe the signature of Pop~III GRBs, and end with a brief outlook into the future in Section~5. In concluding, we would like to refer the reader to two recent reviews, one providing a comprehensive perspective on modern GRB astrophysics \citep{kumar15}, and the other a concise summary of the lessons from the {\it Swift} era for GRB cosmology \citep{salvaterra15}.

\section{Formation of Population~III stars}
\label{sec:1}
The longstanding consensus view has been that the conditions in the early universe favored the formation of predominantly massive stars, such that the Pop~III IMF was top-heavy \citep{Abel02,BCL02,Bromm04}. This expectation rests on the much less efficient cooling in pure H/He gas, where the only viable cooling agent is molecular hydrogen. The primordial gas can therefore reach temperatures of only $\sim 200$~K, compared to the 10~K reached in dust-cooled molecular clouds in the present-day Milky Way. The correspondingly enhanced thermal pressure is reflected in a Jeans mass that is larger by one to two orders of magnitude in the Pop~III case. Another element of this ``standard model'' of primordial star formation has been that the first stars formed typically in isolation, one per minihalo \citep{omukai99}.

In recent years, this traditional paradigm has been refined in important ways \citep{Turk09,Stacy10,Clark11,Greif11,Greif12}. Supercomputing power, as well as algorithmic advances, now enable us to follow the protostellar collapse to densities, $n\sim 10^{22}\;{\rm cm}^{-3}$, where the initial hydrostatic core forms in the center of the cloud \citep{Yoshida08}. Crucially, the computations can now also be extended into the main accretion phase \citep{omukai03}. An important lesson has been that accretion is mediated through a near-Keplerian disk, similar to present-day star formation. The hot conditions in the surrounding cloud result in extremely large rates of infall onto the disk ($\dot{M}\propto T^{3/2}$); this rapid mass-loading drives the disk inevitably towards gravitational instability, such that a small multiple of Pop~III protostars emerges, often dominated by a binary system. It is not yet possible to extend such ab-initio simulations all the way to the completion of the protostellar assembly process; the final mass of Pop~III stars and their final IMF are thus still subject to considerable uncertainty. However, first attempts to carry out the radiation-hydrodynamical calculations required to treat the late accretion phase, where protostellar feedback tends to limit further infall, have confirmed the basic prediction: the first stars were typically massive, with masses of a few $\sim 10 M_{\odot}$, although rarely very massive ($>100 M_{\odot}$), as previously thought, forming as a member of small multiple systems \citep{McKee08,Hosokawa11,Stacy12}. There are indications, though, that the Pop~III mass could occasionally grow to $>300 M_{\odot}$, in cases of unusually weak protostellar feedback \citep{Hirano14,Susa14}.

Are Pop~III stars suitable GRB progenitors? To successfully trigger a collapsar event, the leading contender for long-duration GRBs \citep{Woosley93,MacFadyen01}, a number of conditions have to be met: a central BH has to form, a relativistic jet has to escape the stellar envelope before being quenched, and there has to be a sufficient degree of angular momentum close to the center, to delay the accretion of material onto the BH. These are quite stringent, and often difficult to fulfill simultaneously \citep{Fryer04,Petrovic05,Belczynski07}.

The first requirement for a collapsar central GRB engine, the emergence of BH remnants, is fulfilled because of the top-heavy nature of the primordial IMF. The binary nature of Pop~III stars may allow to meet the second requirement of providing an escape channel for the jet, if the binary is sufficiently close to allow for Roche-lobe overflow and a common-envelope phase, to expel the extended hydrogen (and helium?) envelope. This may be crucial to prevent the quenching of the relativistic jet, launched by the central engine \citep{Bromm06} \citep[but see][and the discussion on this key point below]{Suwa11}. 
Simulations have begun to constrain the statistics of Pop~III binaries,
within a fully cosmological context
\citep{StacyBromm13}. The resulting distribution of semi-major axes
is found to be quite broad, with a peak around $\sim 300$\,AU. Those
simulations, however, cannot yet resolve the formation and evolution of
tight binaries, due to the resolution limit of $\sim 20$\,AU. Such
improved simulations would be required to probe the regime of contact
binaries, where Roche-lobe overflow or common-envelope evolution could
occur during the later red supergiant phase. It appears likely, though,
that such tight binaries exist. A fraction of the sink particles
that numerically represent Pop~III stars in the simulations undergo mergers
when they approach to within the resolution limit. With better resolution,
some of those sinks/stars may well survive as tight binaries.

What about the additional requirement that the collapsar progenitor retains enough angular momentum? This question ties in with the rate of rotation of Pop~III stars, where almost nothing is known yet. A first attempt to address this within a fully cosmological context has recently been carried out \citep{Stacy11,Stacy13}, indicating that the first stars may have typically been very fast rotators, with surface rotation speeds of a few 10 percent of the break-up value. Such high rates of rotation would have important consequences for Pop~III stellar evolution, possibly enabling strong mixing currents, and for the fate encountered at death \citep{Yoon06}. Thus, it is plausible that all requirements for a collapsar central engine were in place in the early universe.

An important caveat here is that any effect of possible magnetic braking on the final
stellar rotation rate has been neglected so far. Recent studies have
argued that turbulent dynamo amplification in the primordial protostellar disks
might rapidly build up dynamically significant magnetic fields \citep[e.g.,][]{schober12}.
It is currently not clear whether such small-scale, tangled fields could
be organized into larger scale arrangements. If they can, magnetic torques
may be responsible for establishing Pop~III stellar rotation rates similar
to what is known for present-day O~stars. Ongoing magneto-hydrodynamical
simulations should soon help to clarify this key point.

\section{Evolution of Pop III stars and GRBs}

Metallicity is one of the prime factors that determine the evolution of massive
stars. Many features of stellar evolution are therefore uniquely found with
massive Pop III stars, compared to the case of metal-rich counterparts.

Massive stars on the main sequence in the nearby universe
are powered by the CNO cycle.  In the early universe, CNO elements
are absent and core hydrogen burning  starts with the pp chain. Because energy
production by the pp-chain is too weak to maintain thermal equilibrium,  a
massive Pop III star initially undergoes thermal contraction until the central
temperature becomes high enough for helium burning to produce carbon. The
CNO cycle then becomes active with thus-produced carbon, and the structure of
the star is adjusted accordingly until thermal equilibrium is reached. The main
sequence evolution begins only thereafter~\citep[e.g.,][]{Marigo01, Ekstroem08,
Yoon12}. 

The evolution on the main sequence and in later stages is more critically
affected by the lack of heavy elements. In metal-rich environments, massive
star evolution is characterized  by strong mass loss by radiation-driven winds,
for which iron lines play a particularly important
role~\citep[e.g.,][]{Puls08}.  For massive Pop III stars, the radiation force
resulting from hydrogen and helium lines is too weak to drive a wind, and the
predicted mass loss rate is extremely low (i.e., $\dot{M} \le
10^{-14}~\mathrm{M_\odot~yr^{-1}}$; \cite{Krticka06}). 

\begin{figure}
\includegraphics[width=\textwidth]{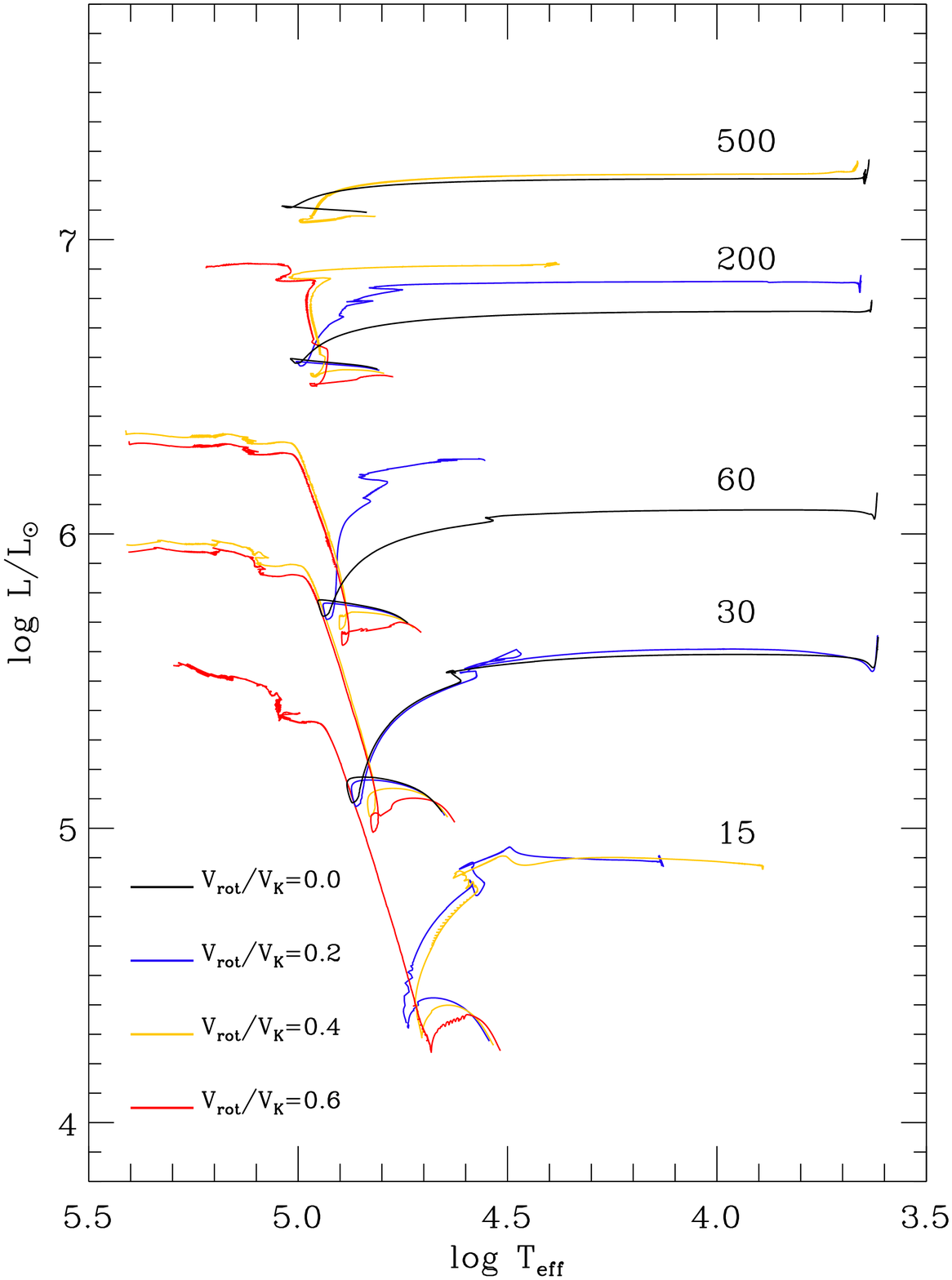}
\caption{Evolution of massive Pop III stars (15, 30, 60, 200 and 500~$\mathrm{M_\odot}$) with and without rotation. 
The adopted initial rotational speeds in units of the Keplerian value are marked by different colors as indicated by the labels.
These evolutionary models are taken from \citet{Yoon12}.} 
\label{fig:hr}       
\end{figure}

This has two important consequences. First, massive Pop III stars would not
easily lose their initial angular momentum via mass loss. This implies that the
evolution of massive Pop III stars could be dominated by the effects of
rotation (Sects.~\ref{sect:rotation}~and~\ref{sect:che}). Second,  even very
massive stars ($M \ga 100~\mathrm{M_\odot}$) that are close to the Eddington
limit would retain their hydrogen envelopes until the end of the evolution,
This can produce  very massive red supergiant stars (see Fig.~\ref{fig:hr}),
which are not found in the local universe.  Metallicity effects also critically
influence the condition for convection inside stars, which  plays an important
role in the final structure at the pre-collapse stage (Sect.~\ref{sect:blue}).
All of these effects are closely related to the properties of Pop III GRB
progenitors, as discussed below. 

\subsection{The role of rotation}\label{sect:rotation}

\subsubsection{Evolution of angular momentum}

One of the key necessary conditions for GRB progenitors is rapid rotation. To
trigger relativistic jets, a large amount of  angular momentum  should be
retained in the innermost layers of GRB progenitors until the pre-collapse
stage. More specifically, the specific angular momentum in the core must be higher than   $j > 1.5
\times 10^{16}\left(\frac{M_\mathrm{BH}}{3~\mathrm{M_\odot}}\right)
~\mathrm{cm^2~s^{-1}}$ within the collapsar scenario, where $M_\mathrm{BH}$ is
the black hole mass \citep{Woosley93}, and $ j \simeq
4\times10^{15}~\mathrm{cm^2~s^{-1}}$ within the magnetar scenario
~\citep[e.g.,][]{Wheeler00}.

Observations indicate that the amounts of angular momentum are sufficient to
meet these criteria for a large fraction of massive main-sequence stars in the
nearby universe~\citep[e.g.,][]{Mokiem06, Wolff06, Ramirez13}. Recent numerical
studies also indicate that massive Pop III stars would be rapid
rotators~\citep{Stacy11, Rosen12}. However, numerous observations imply rapid
transport of angular momentum inside stars~\citep[e.g.,][]{Suijs08,
Charpinet09, Eggenberger12, Marques13}, which must have significant impact on
the final angular momentum distribution in the stellar core~\citep{Hirschi04,
Heger05}. While the main mechanism for angular momentum transport is still a
matter of debate, the following two mechanisms have been considered most
important in stellar evolution models.  

\begin{itemize}
\item \emph{Eddington-Sweet circulations:}
Thermal equilibrium in rotating stars generally breaks down because the
radiative flux along the polar axis becomes excessive compared to that along
the equatorial one (the von Zeipel theorem; e.g., \citealt{Kippenhahn90}).  To
compensate this thermal imbalance, large-scale meridional circulations are
induced in rotating stars, which are nowadays often called `Eddington-Sweet
(ES) circulations'~\citep[e.g.,][]{Maeder12}. These ES circulations are not
only important for the energy flux, but also for the transport of angular
momentum.  The timescale for ES circulations in a chemically homogeneous star
is roughly given by $\tau_\mathrm{ES} \approx
\tau_\mathrm{KH}(\Omega/\Omega_\mathrm{K})^{-2}$, where $\tau_\mathrm{KH}$ denotes
the Kelvin-Helmholtz timescale and $\Omega/\Omega_\mathrm{K}$ the ratio of the
angular velocity  to the Keplerian value. Although  $\tau_\mathrm{ES}$ can be
in principle shorter than the nuclear timescale ($\tau_\mathrm{nuc}$) of a
star,  ES circulations are severely slowed down once the chemical
stratification across the boundary between the core and the hydrogen envelope
is built up~\citep{Maeder98}.  Therefore, it is believed that ES circulations
cannot efficiently brake down the stellar core for most cases~\citep{Heger00,
Hirschi04}.  
\item \emph{Taylor-Spruit dynamo:}  According to \citet{Taylor73},
toroidal fields in radiative layers of stars are always unstable to create
poloidal fields (see also \citealt{Spruit99}).  If differential rotation can
wind up thus-created poloidal fields to amplify toroidal fields, the dynamo
loop can be closed~\citep{Spruit02}.  This so-called Taylor-Spruit (TS) dynamo may lead to
magnetic torques that can redistribute angular momentum inside stars much more
efficiently than ES circulations. Theoretical studies indeed show that
solid-body rotation can be maintained in main sequence stars with TS dynamo~\citep{Heger05}.  
This rapid
angular momentum transfer by the TS dynamo has been invoked to explain the
relatively slow spin rates of young neutron stars and isolated white dwarfs, as
well as angular velocity profiles in low-mass stars on various evolutionary
stages that are inferred from asteroseismological data~\citep[e.g.][]{Eggenberger05, Heger05, Suijs08, Cantiello14}.  
However, the validity of the TS dynamo theory  is still debated~\citep[e.g.,][]{Braithwaite06, Zahn07, Arlt11, Cantiello14}. 
\end{itemize}

Evidently, the transport mechanisms may not be limited to these ones.
Rotationally-induced hydrodynamic instabilities apart from ES circulations
include the shear instability and the baroclinic instability.   The interplay
between ES circulations and the shear instability would be particularly important
for the transport process~\citep{Maeder12}.  Several authors also investigated
the role of the magneto-rotational instability~\citep{Wheeler15} and internal
gravity waves in massive stars~\citep{Fuller14}.  More progress is certainly
needed to have a reliable prescription for angular momentum transport. 

\subsubsection{Chemical mixing}

Rotationally-induced hydrodynamic instabilities can transport not only angular
momentum but also chemical species.  This may lead to chemical mixing of
hydrogen-burning products from the convective core into the radiative envelope in
a massive star on the main sequence.  Enhanced abundances of nitrogen and
helium  are indeed found at the surfaces of many massive stars, which provide
evidence for such mixing~\citep[e.g.][]{Hunter08, Hunter09}. 
The efficiency of chemical mixing due to rotation has been
recently calibrated by \citet{Brott11} using the inferred nitrogen abundances
at the surfaces of B-type stars in the Large Magellanic Cloud. 
But the anomalously high/low nitrogen abundances observed in some
slowly/rapidly rotating stars have  not been well understood
yet~\citep{Hunter08, Hunter09, Brott11, Aerts14}.

\subsubsection{Mass shedding}

Rapidly rotating stars can reach the critical rotation during the course of
their evolution if they do not lose a sufficient amount of angular momentum via
strong mass loss. Note that the critical rotation speed ($v_\mathrm{crit}$) can 
become lower than the Keplerian limit ($v_\mathrm{K} = \sqrt{GM/R}$)
if the stellar luminosity approaches the Eddington limit, as the following: 
\begin{equation}
v_\mathrm{crit} = v_\mathrm{K}\sqrt{1 - \Gamma}~,  
\end{equation}\label{eqvcrit}
where $\Gamma$ is the Eddington factor~(\citealt{Heger00}; see \citealt{Maeder00} for an
alternative description for the critical rotation speed). 
Once the star reaches the critical rotation, mechanical mass shedding would occur, even
when radiation-driven winds are very weak. 
This would be the dominant mode of mass-loss from massive Pop III stars
~\citep{Marigo03, Ekstroem08, Yoon12}.  
For very massive stars ($M \ga 100~\mathrm{M_\odot}$), however, pulsationally driven winds 
during the red supergiant phase might also play an important role~\citep{Baraffe01, Moriya15}.

\subsection{Chemically homogeneous evolution and GRB progenitors}\label{sect:che}

Typical timescales about $\sim$ 10 secs of long GRBs imply that their
progenitors are generally compact ($R \la 10~\mathrm{R_\odot}$;
\citealt{Woosley06}).  All of the supernovae associated with a long GRB turn out to
be Type Ic, which provide further evidence for compact progenitors that have lost their
hydrogen envelopes~\citep[e.g.,][]{Woosley06, DElia15}.  It is therefore widely
believed that the majority of long GRB progenitors are naked helium stars like Wolf-Rayet
(WR) stars.  Another necessary condition for long GRB progenitors is rapid
rotation as mentioned above. 

Mass loss by radiation-driven winds is negligible for Pop III stars, and mass
shedding due to rotation is not significant enough to remove the whole hydrogen
envelope~\citep{Marigo03, Ekstroem08, Yoon12}. This may raise a question of
whether massive single Pop III stars can produce an ordinary long GRB.  

Probably, the only possible solution for ordinary long GRBs from single Pop III
stars would involve the chemically homogeneous evolution (CHE).  CHE may occur with
a sufficiently high rotation speed if the timescale for chemical mixing due
to ES circulations
becomes shorter than the nuclear timescale~\citep{Maeder87}.  In other words,
CHE can be realized if rapid chemical mixing between the hydrogen-burning
convective core and the radiative envelope occurs, before nuclear burning
builds up a strong degree of chemical stratification across the boundary
between them that would dramatically slow down ES circulations. In this case,
almost all the hydrogen in the envelope of a star is mixed into the
hydrogen-burning core to be fused into helium, which is in turn mixed back into
the envelope.  The whole star is thus gradually transformed into a helium star
by the end of the main sequence. This makes the CHE stars evolve blueward as
shown in Fig.~\ref{fig:hr}.  

\begin{figure*}
\includegraphics[width=0.47\textwidth]{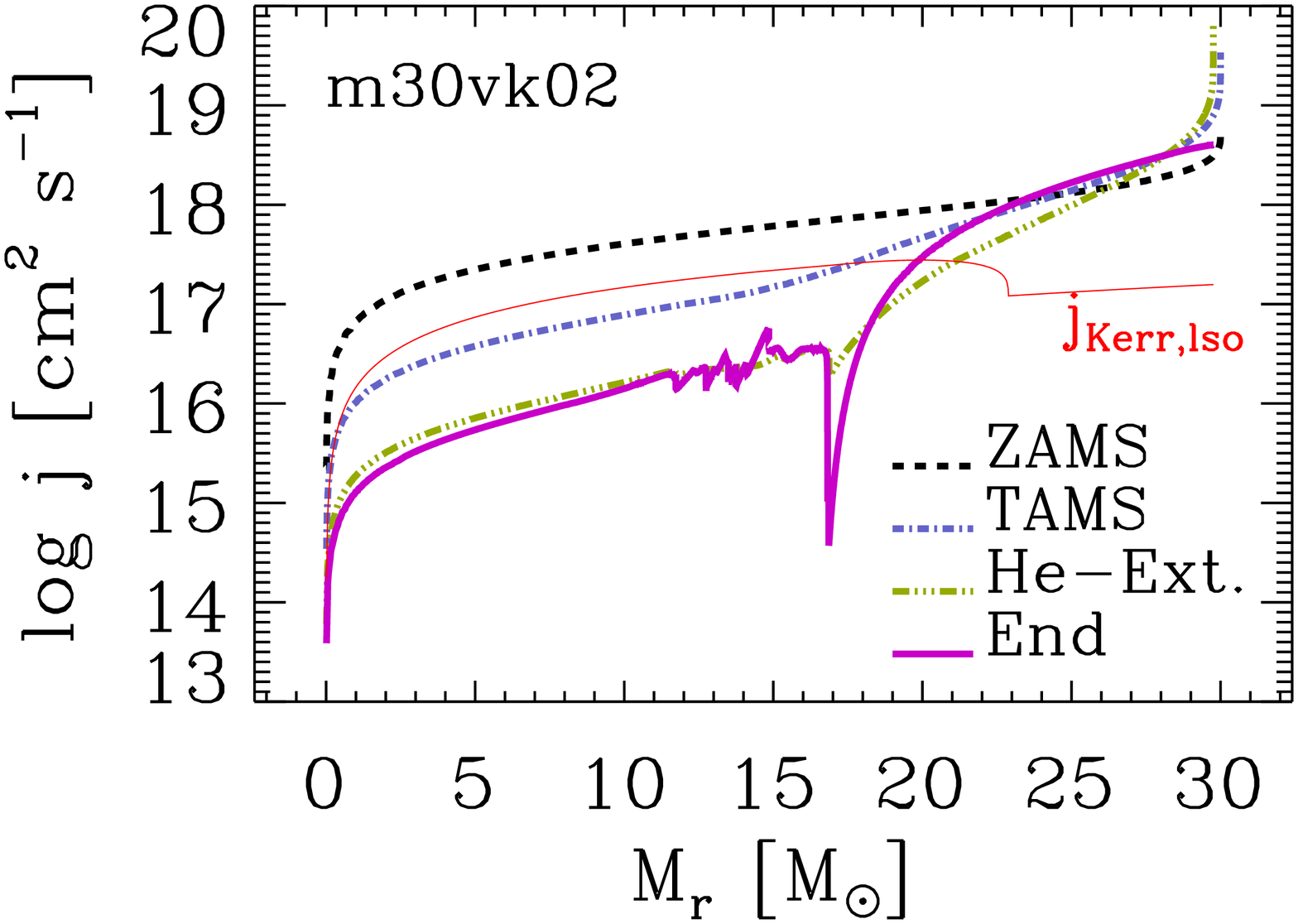}
\includegraphics[width=0.47\textwidth]{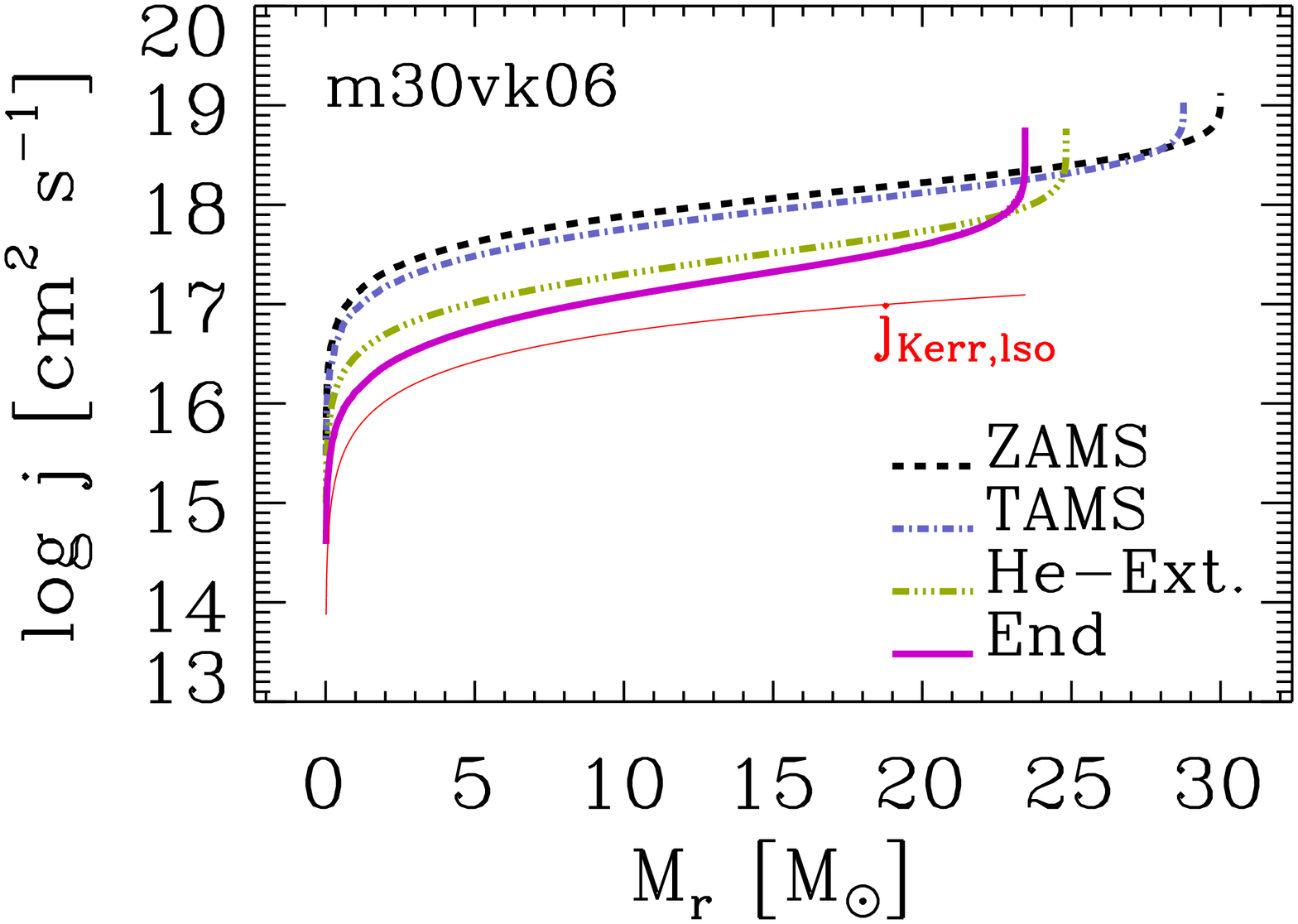}
\caption{Mean specific angular momentum profile as a function of the mass coordinate for
different evolutionary epochs of 30~$\mathrm{M_\odot}$ Pop III star: zero age main sequence (ZAMS), terminal age of the main sequence (TAMS), 
core helium exhaustion (He-Ext.) and the last calculated model, which corresponds to core neon exhaustion. 
The TS dynamo was included in the calculations, and the two different initial rotational speeds were adopted: 20\% and 60\% Keplerian
rotation speeds for the left and the right panels, respectively. In the latter case, the star evolves chemically homogeneously.
Adapted from \citet{Yoon12}.}
\label{fig:jspec}       
\end{figure*}

In the case of normal evolution, the helium core during the post-main sequence
stage would be significantly braked down by the slowly rotating hydrogen
envelope that expands to a red-supergiant phase, which can serve as a large angular
momentum reservoir~\citep{Heger05}. By contrast, CHE stars do not undergo the
red-supergiant phase and can retain a large fraction of the initial amount of
angular momentum until the pre-collapse stage~(Fig.~\ref{fig:jspec};
\citealt{Yoon05, Woosley06b}). CHE can be more easily realized at lower
metallicity~\citep{Yoon06}, and therefore CHE is a promising pathway to long
GRBs for massive Pop III stars ~\citep{Yoon12}. 

\begin{figure*}
\includegraphics[width=\textwidth]{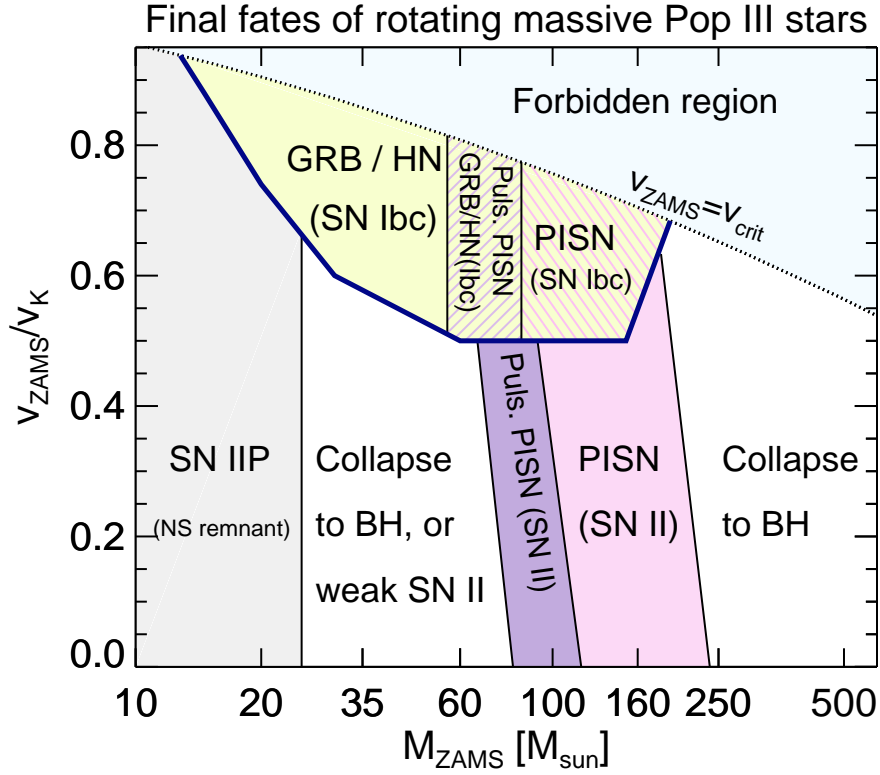}
\caption{Final fates of rotating massive Pop III stars, in terms of the initial mass and the initial rotational speed in units of the Keplerian value.
The boundary for the CHE regime is marked by the thick solid line. The dotted line denotes the critical rotational speed for a given ZAMS mass, 
as defined by Eq.~(\ref{eqvcrit}). This figure is a reproduction of Figure 12 in \citet{Yoon12} with permission from Astronomy \& Astrophysics, \copyright~ESO.}
\label{fig:final}       
\end{figure*}

As summarized in Fig.~\ref{fig:final}~\citep[see also][]{Yoon12}, CHE would
produce GRB progenitor candidates for the initial mass range of 13 -
84~$\mathrm{M_\odot}$ if the initial rotation speed is higher than about 50\%
of the Keplerian value.  Another interesting prediction of CHE is that the
initial mass range for pair-instability supernovae (PISNe) would significantly
decrease to 84 --- 190~$\mathrm{M_\odot}$, compared to the non-rotating case
(i.e., 140 - 300~$\mathrm{M_\odot}$; see also \citealt{Chatzopoulos12}).  These
PISNe from CHE stars would appear as Type Ic, instead of Type IIP that is
expected for non-rotating or slowly rotating PISN progenitors.  For the initial
mass range of 56-84~$\mathrm{M_\odot}$ of GRB progenitors, a pulsational PISN
would occur shortly before the core collapses into a black hole, which may
produce a very bright optical event via an interaction between the pulsational
PISN ejecta and the GRB jet that follows.

\subsection{Supergiant Pop III  progenitors for ultra-long GRBs}\label{sect:blue}

With relatively slow rotation,  the typical core-envelope structure is
developed in a massive Pop III star after the end of core hydrogen burning.
The helium core is expected to be spun down as the core angular momentum is
transferred to the expanding hydrogen envelope.  At the pre-collapse stage, the
inner core would rotate relatively slowly and its specific angular momentum
would not exceed the critical limit for producing a GRB. 

Interestingly, because Pop III stars do not lose much mass, the total angular
momentum can be more or less conserved and the specific angular momentum of the
hydrogen envelope increases as a result of  the angular momentum transfer from
the core (see the left panel of Fig~\ref{fig:jspec}).  In this case, 
the core would directly collapse into a BH, and the envelope material would 
form a Keplerian disk around it.  The consequent accretion rate
depends on the free-fall time of the envelope material: it would be $\sim
10^{-4}~\mathrm{M_\odot~yr^{-1}}$ for a blue-supergiant (BSG) progenitor, and
$\sim 10^{-6}~\mathrm{M_\odot~yr^{-1}}$ for a red-supergiant (RSG)
progenitor~\citep{Woosley12}.  Therefore, supergiant Pop III stars could
produce an ultra-long GRB having a timescale of $10^3 - 10^7$~sec
\citep[cf.][]{Gendre13, Virgili13, Evans14, Levan14, Piro14}, depending on the
final structure~\citep{Suwa11, Nakauchi12, Woosley12, Yoon15}. 

Note, however,  that the hydrogen envelope of a RSG  is very loosely bound.
Recently, \citet{Lovegrove13} showed that most of the RSG envelope may be ejected
as a result of the rapid loss of gravitational mass due to neutrino emission from the core while a BH is
formed. This makes it difficult for a RSG to produce a long GRB
transient even if its envelope is rotating rapidly.  By contrast, the binding
energy of a BSG envelope is fairly high and it would remain gravitationally
bound to the newly formed BH.  Therefore, rapidly rotating BSGs should be
considered good progenitor candidates for ultra-long GRBs with a timescale of
about $10^3 - 10^4$ sec~\citep{Woosley12}. 

The remaining question is whether massive Pop III BH progenitors end their
lives as a BSG or a RSG. Although the relatively low-opacity with zero-metallicity
helps a massive Pop III star remain compact, non-linear effects of stellar
evolution make it very difficult to make a robust prediction on the final
structure.  For example, lower opacity in the core of a Pop III star than in a
metal-rich counterpart leads to a smaller size of the convective core on the
main sequence for a given mass, which in turn results in a more compact
carbon-oxygen core at the end of core helium burning. As a consequence, the
helium-burning shell source becomes very hot, inducing a more violent convection
above it. This often leads to penetration of the convection zone of the helium shell source into the
hydrogen-burning shell, boosting the CNO cycle~\citep{Heger10, Limongi12,
Yoon12}. The huge amount of nuclear energy produced in this way can make the star
become a RSG shortly before core collapse.  The onset of this CNO boosting
sensitively depends on the adopted overshooting parameter and the initial
rotation speed, but it is usually found for Pop III stars in the mass range of
about 10 - 30~$\mathrm{M_\odot}$~(C. Kye \& S.-C. Yoon in prep).  For more
massive Pop III stars, the RSG solution seems to be generally
favored~\citep{Marigo01, Marigo03, Ekstroem08, Yoon12}, but for a limited mass
range of about $30 - 60~\mathrm{M_\odot}$, the BSG solution could be obtained
with moderate rotation or weak overshooting (\citealt{Limongi12, Yoon12}; C. Kye \&
S.-C. Yoon in prep; see also Fig.~\ref{fig:hr}).

\subsection{Very massive Pop III stars and supercollapsars}\label{sect:vms}

For the mass range above the PISN limit  and below the general relativistic
instability limit (e.g., $ 260~\mathrm{M_\odot} \la M \la
50000~\mathrm{M_\odot}$ without rotation),  Pop III stars would directly
collapse to a BH with $M \ga 100~\mathrm{M_\odot}$ \citep[e.g.,][]{Fryer01}.
This would lead to a collapsar that may produce a long GRB having a very
high total energy ($\ga 10^{53} - 10^{54}~\mathrm{erg}$) and a very long
timescale ($T \sim 10^4 - 10^6\mathrm{sec}$), depending on the final structure of
the progenitor star (the so-called supercollapsar; \citealt{Komissarov10, Meszaros10}). 

In terms of stellar evolution, the following issues should be considered to
probe the possibility of Pop III supercollapsars. 

\begin{itemize}
\item Very massive stars are close to the Eddington limit, and can reach the critical
rotation at a relatively low rotation speed. Angular momentum loss through mass
shedding  becomes significant accordingly, which slows down ES circulations.
For this reason, CHE cannot be easily realized for very massive Pop III stars with
$M \ga 200~\mathrm{M_\odot}$ \citep{Yoon12}. This means that they would not be able to avoid
developing the typical core-envelope structure during the post-main sequence phase. 
\item The angular momentum transport from the core to the envelope can easily 
remove the core angular momentum in this case, if there exists a strong coupling
between them via, for example, magnetic torques. The angular momentum condition 
for a collapsar cannot be fulfilled in this case~\citep{Yoon12}. By contrast, if ES circulations 
were the dominant mode of angular momentum transfer, 
the core and/or the envelope would retain a sufficient amount of angular momentum
to produce a collapsar for a limited mass range of about $300~\mathrm{M_\odot} \la M
\la 700~\mathrm{M_\odot}$~\citep{Yoon15}. 
\item The RSG solution is strongly favored for these very massive stars:
it is most likely that Pop III stars with $M > 260~\mathrm{M_\odot}$ end their lives as a RSG~\citep{Marigo03, Yoon12, Yoon15}. 
Therefore, even if the core was rotating rapidly enough to produce a collapsar, 
the resultant relativistic jet would not be able to penetrate the envelope. 
The final outcome would be a jet-driven Type IIP supernova, rather than a GRB. 
The caveat here is that  the mass loss rate from such a very massive RSG 
is not well understood.  RSG stars are usually unstable to radial pulsations, and pulsation-driven
winds might be significant. Whether or not such winds can completely remove the hydrogen
envelope of a very massive Pop III RSG star is  a matter of debate \citep[e.g.,][]{Baraffe01, Moriya15}.  
\item 
For a supercollapsar, a neutrino-driven jet cannot be efficient because of the large
radius of the last stable orbit around such a massive black hole.  Therefore,
another important ingredient for a successful supercollapsar is strong
large-scale magnetic fields in the progenitor star such that a relativistic
jet may be triggered via the Blandford-Znajek (BZ) mechanism~\citep[][see also Section~4]{Komissarov10, Meszaros10}. 
Magnetic field configuration in massive stars  
at the pre-collapse stage has been hardly investigated so far, and we still do not know 
if this condition of strong large-scale magnetic fields can be fulfilled in very massive Pop III stars. 
\end{itemize}

In short, a supercollapsar seems to be rather difficult to produce from very
massive Pop III single stars, given that they would not follow CHE and that
they would become a RSG at the pre-collapse stage, rather than a BSG or a naked
He star.  However, binary interactions might produce a BSG supercollapsar
progenitor more easily, as discussed in Sect.~\ref{sect:binary} below.

\subsection{Binary interactions and GRB progenitors}\label{sect:binary}

Binary interactions would be an alternative way to produce naked helium stars
at zero metallicity.  However, it is expected that they are slow rotators  in
general and would not produce a GRB~\citep{Yoon10}: GRB production even from
binary stars would require an exotic evolutionary path.  Several different
scenarios for binary GRB progenitors have been suggested~\citep[see][]{Brown00,
Izzard04, Fryer05, Heuvel07, Podsiadlowski10}, but because of the complexity of
the related physical processes like common envelope ejection, the details of
these scenarios have not been properly investigated with self-consistent binary
evolution models for most cases. 

Here we summarize the possibly important aspects of binary stars for Pop III GRBs, 
which should be carefully studied in the near future. 
\begin{itemize}
\item Mass exchange in close binary systems can enhance
the number fraction of rapidly rotating stars~\citep{deMink13}. 
As shown by \citet{Cantiello07}, the Case B mass transfer (i.e., mass transfer
during the helium core contraction phase or the early stage of core helium burning) is likely to enhance the possibility 
for CHE in mass accretors. The parameter space for this solution should be systematically explored for Pop III binaries. 
\item In the literature,  
evolutionary channels including the Case C mass transfer (i.e, mass transfer during the late stage of core helium burning or thereafter) has been often 
invoked to explain GRB progenitors~\citep{Brown00, Heuvel07, Podsiadlowski10}. In metal-rich environments, strong mass loss tends to prohibit the Case C mass transfer
from massive stars with $M > 30~\mathrm{M_\odot}$ that are potential BH progenitors, but it would occur more frequently for Pop III stars
that do not lose much mass. It is therefore possible that the Case C channel for long GRBs  would be more important for Pop III stars than 
in the nearby universe. 
\item Binary mergers (in particular Case B mergers) may produce a BSG progenitor more easily, 
because a fairly small mass ratio of the helium core to the hydrogen envelope can be achieved in the merger remnants~\citep[e.g.,][]{Justham14}. 
Although the helium core may not be necessarily rapidly rotating in this case, 
the hydrogen envelope must have a very high specific angular momentum until the pre-collapse stage
because mass loss from Pop III stars is weak.  
Therefore, Case B mergers at low-metallicity would be a very promising  pathway to ultra-long GRBs. 
It would be particularly interesting if very massive BSG GRB progenitors ($M \ga 300~\mathrm{M_\odot}$)
could be produced in this way~\citep[see][for a more detailed discussion]{Yoon15}. 
\end{itemize}

\section{Observational signatures in Pop III GRBs}

\subsection{How to identify Pop III GRBs?}

Recent studies on stellar formation in the early universe predict that Pop III stars may be most prominent at $z \sim 20-30$ \citep[e.g.][and see Sections 1 and 2]{Abel02,Yoshida08,Bromm11}. Observations of GRBs may provide unique probes of the physical conditions of the universe at such redshifts. Due to their high luminosities, GRB prompt emissions and afterglows are expected to be observable at least out to $z \ga 10$ \citep[e.g.][]{Lamb00,Ciardi00,Bromm02,yonetoku04,Gou04}. This can serve as a tracer of the history of the cosmic star formation rate \citep{totani97,porciani01,Bromm06,kistler09,desouza11} and provide invaluable information on the physical conditions in the intergalactic medium \citep{barkana04,ioka05,Inoue07}. Currently, the most distant GRB that has been spectroscopically confirmed is GRB 090423 at $z \simeq 8.2$ \citep{Tanvir09,Salvaterra09}, and GRB 090429B has a photometric redshift $z \simeq 9.4$ \citep{cucchiara11}. The detailed spectroscopic observation of GRB 050904 at $z \simeq 6.3$ has put a unique upper bound on the neutral hydrogen fraction in the intergalactic medium at that redshift and estimated the metallicity \citep{Totani06,kawai06}. See also a recent detailed analysis of GRB 130606A at $z \simeq 5.9$ and related debates \citep[][and references therein]{totani15}. These observations indicate that GRBs are very promising for exploring the high-redshift universe \citep[for a recent review, see][]{salvaterra15}. 

In the context of the study of Pop III stars, a crucial question is how can we identify GRBs originating from Pop III stars (Pop III GRBs) among bursts whose redshifts are determined as $z \ga 10$ e.g. by observation of the Ly$\alpha$ cutoff at the IR frequencies. An unambiguous way to pinpoint a Pop III progenitor is to examine whether the afterglow spectrum from its surrounding medium is devoid of iron-group elements\footnote{It may be misleading to state ``devoid of metals''. Pop~III GRB progenitors via CHE, for example, could have ejected CNO elements into the surrounding medium via mass shedding during the late stages of their evolution.} through high-resolution IR and X-ray spectroscopy by ground-based facilities and/or future space experiments. However, there may also be the case that the surrounding medium is embedded in a region where stellar explosions have already occurred slightly earlier, and the absorption lines of the first heavy elements produced by those explosions are thus imprinted \citep{Wang12}. A true Pop III origin of the GRB progenitor could then be missed due to this masquerading effect. It is therefore important to explore alternative strategies to identify Pop III GRBs. In this regard, another proposed way is to focus on the total energies and durations of GRBs estimated by the X-ray and gamma-ray observations of prompt emissions and/or afterglows \citep{Komissarov10,Meszaros10,toma11,Suwa11}. Pop III stars could be very massive stars (VMSs) with $M \ga 300 M_\odot$, and then GRBs originating from Pop III VMSs (Pop III VMS GRBs) could have peculiarly large total energies and long durations.\footnote{Recently, GRBs from supermassive stars with $M \sim 10^5 M_\odot$ which could make seed BHs for supermassive BHs are also discussed \citep{matsumoto15}.} Their values could be orders of magnitude larger than those of ordinary Pop I and Pop II GRBs, providing strong hints for Pop III progenitors. 

The recent numerical studies of the evolutions of Pop III stars imply that luminous GRBs are rather difficult to occur from VMSs \citep{Yoon12,Yoon15}, as discussed in Section 3. However, it may not be interpreted as a robust prediction yet, since theoretical modelings of formation and evolution of Pop III stars are highly non-linear, complex problems, as also noted in Section 3. There are several factors to be investigated, such as the multi-dimensional effects related to magnetic field and stellar rotation, the radiative feedback, the mass accretion after the onset of core hydrogen burning, the binary interactions, and so on. If only Pop III stars with $M < 100 M_\odot$ could produce GRBs, it would not be simple to identify Pop III GRBs through the estimate of the total energy and duration scales. Such bursts at the high redshifts may not be detectable by the current satellites \citep{Nakauchi12}, and then persistent efforts on observational and theoretical studies with new high-sensitivity satellites would be needed to find out possible statistical differences between Pop III GRBs and ordinary GRBs. On the other hand, the history of GRB study tells us that some breakthroughs happened by simple observations, such as the serendipitous discovery of GRBs itself, the isotropic angular distribution of GRBs in the sky revealed by BATSE, the discovery of the afterglow by Beppo-SAX, the confirmation of the connection between GRBs and supernovae by HETE-2, and several recent discoveries by Swift and Fermi \citep[for reviews, see e.g.][]{zhang04,kumar15}. The connection between GRBs and Pop III stars might also be discovered by a serendipitous observation of a source which looks peculiar. Here we consider a case in which Pop III VMSs can produce GRBs as luminous as ordinary GRBs and discuss their unique observational signatures. Searches of such peculiar GRBs will constrain the theoretical models of the formation and evolution of Pop III stars.

\subsection{Pop III VMS GRB model}
\label{subsec:GRBmodel}

Let us first make a phenomenological interpretation of the total gamma-ray energies and durations of ordinary long GRBs which are currently observed. Focusing on the most energetic GRBs, which are mainly detected by Fermi satellite, they have $E_{\gamma,iso} \sim 10^{54} - 10^{55}\;$erg. The typical duration is $T \sim 10 - 100\;$s \citep{ackermann13}. In the collapsar model, the total gamma-ray energy can be written as
\begin{equation}
  E_{\gamma,iso} = \frac{\epsilon_{\gamma} \eta M_{\rm acc} c^2}{1- \cos\theta_j} \sim 6 \times 10^{54}\; \epsilon_{\gamma} \left(\frac{\eta}{10^{-2}}\right) \left(\frac{M_{\rm acc}}{M_\odot}\right) \left(\frac{\theta_j}{0.1}\right)^{-2}\;{\rm erg},
\label{eq:Eiso}
\end{equation}
where $\epsilon_{\gamma}$ is the gamma-ray radiation efficiency, $\eta$ the conversion factor from the accretion energy to the jet energy, $M_{\rm acc}$ the total mass accreted by the central black hole (BH), and $\theta_j$ the opening half angle of the jet. While $\epsilon_{\gamma} \ga 0.5$ and $\theta_j \sim 0.1$ are supported by some observational indications \citep{ioka06,zhang07,frail01} \citep[but see][]{beniamini15}, the values of $\eta$ and $M_{\rm acc}$ are highly uncertain. (Furthermore, the conversion factor from the accretion luminosity $\dot{M}_{\rm acc} c^2$ to the jet luminosity may be time-dependent, so that $\eta$ in equation (\ref{eq:Eiso}) should be interpreted as a rough temporally-averaged value of the energy conversion factor.) In a scenario in which the progenitor star has an initial mass $\sim 30\; M_\odot$, ends its life as a Wolf-Rayet star of $\sim 15\;M_\odot$ after losing a large fraction of mass by the stellar wind, and collapses making a BH of $\sim 3\;M_\odot$, being accompanied by a supernova explosion with an ejecta mass of $\sim 10\;M_\odot$ \citep{iwamoto98,mazzali06}, the total accreted mass is $M_{\rm acc} \sim M_\odot$. If the stellar wind is so weak that the star does not lose much of its mass, we have $M_{\rm acc} \sim 10\;M_{\odot}$. Correspondingly, $\eta$ can be estimated to be $\sim 10^{-4} - 10^{-2}$ for $E_{\gamma,iso} \sim 10^{54} - 10^{55}\;$erg.

The conversion factor $\eta$ should be determined by the jet production mechanism. For driving relativistic jets, the thermal energy injection by $\nu \bar{\nu}$ annihilation \citep{eichler89} and/or the electromagnetic energy injection by magnetic braking of the accretion disk or the BH may be viable. In the former thermal model, $\eta$ is estimated to be $\sim 10^{-4} - 10^{-2}$ for $\dot{M}_{\rm acc} \sim (0.1 - 1) \times \;M_{\odot}\;{\rm s}^{-1}$ and $M_{\rm BH} = 3\;M_\odot$ \citep{zalamea11}. However, it depends on the condition of the BH accretion such as the strength of the disk wind \citep[see also][]{suwa13}. In the latter magnetic model, BZ process, i.e. the magnetic braking of the BH or the energy extraction of the BH rotational energy \citep{blandford77}, can give rise to even $\eta > 1$ as demonstrated by MHD numerical simulations \citep{tchekhovskoy11}. There have been some observational indications supporting such a high value of $\eta$ in relativistic jets of active galactic nuclei \citep[e.g.][]{fernandes11,ghisellini14}. In this model also, $\eta$ depends on the condition of the BH accretion as well as on the behavior of the magnetic field \citep[e.g.][]{tchekhovskoy15}. There are also some issues relating to the physics of the BZ process itself \citep{komissarov09,toma14}.

The total duration scale may be estimated as the disk lifetime. If the progenitor star is rotating at half of the break-up speed, then the initial outer edge of the disk is at $R_{\rm d} \simeq R/4$, where $R$ is the stellar radius. The disk lifetime is given by its viscous accretion timescale $\sim R_{\rm d}^2/(\alpha c_s H)$, where $\alpha$ is the effective viscous stress parameter of the $\alpha$-disk model \citep{shakura73}, $c_s$ the sound speed, and $H$ the scale height of the disk. The balance equation between the thermal pressure gradient and the gravitational force gives $c_s/H \sim \Omega_K$, where $\Omega_K$ is the Keplerian angular velocity, so that we have in the thick disk case ($H \sim R_{\rm d}$)
\begin{equation}
  T \sim \frac{1}{\alpha \Omega_K} (1+z) \simeq 200\;\left(\frac{\alpha}{0.1}\right)^{-1} \left(\frac{R}{10^{10}\;{\rm cm}}\right)^{\frac{3}{2}} \left(\frac{M_{\rm BH}}{3 M_{\odot}}\right)^{-\frac{1}{2}} \left(\frac{1+z}{3}\right)\;{\rm s}.
\label{eq:duration}
\end{equation}
We should note that the duration of the bright part of the prompt emission may be shorter than this estimate since the conversion factor from the accretion luminosity to the jet luminosity and the accretion luminosity itself are time-dependent \citep{Komissarov10,kumar08,tchekhovskoy15}.\footnote{\cite{mizuta13} propose an alternative scenario that the duration of the bright part of the prompt emission is determined by an increase of $\theta_j$ on a timescale of the pressure decay of the cocoon emerging from the progenitor star.}

We apply the above scalings of the total gamma-ray energy and duration to a VMS progenitor with $M_* \sim 10^3\;M_\odot$, although those scalings include many uncertainties. If we substitute $M_{\rm acc} \sim 300\;M_\odot$, $M_{\rm BH} \sim 300\; M_{\odot}$, $R \sim 10^{12}\;$cm, and $z \sim 20$ for equations (\ref{eq:Eiso}) and (\ref{eq:duration}), we obtain
\begin{equation}
  E_{\gamma,iso} \sim 10^{57} \left(\frac{\eta}{10^{-2}}\right)\; {\rm erg}, ~~~ T \sim 1\; {\rm day}.
\end{equation}
Therefore, if we were to observe a burst at redshift $z \ga 10$ with such large $E_{\gamma,iso}$ and $T$, this would very likely be a Pop III VMS GRB \citep{Meszaros10,toma11}.

For such large BH masses, the density and temperature of the accretion disk are too low for neutrino cooling to be important, and then the thermal energy from the $\nu \bar{\nu}$ annihilation is insufficient to power strong jets \citep{fryer01,Komissarov10,zalamea11}. The electromagnetic effects such as BZ process may instead power the jets, which should be dominated by Poynting flux \citep[see also][]{suwa07}.

Pop III VMS GRBs share the property of the extremely long durations with Ultra-long GRBs, which were recently detected by Swift with low redshifts \citep[e.g.][]{Gendre13,Levan14}. Ultra-long GRBs have durations as long as $T \sim 10^4\;$s, although their total gamma-ray energies are $E_{\gamma,iso} \sim 10^{53} - 10^{54}\;$erg, which are comparable to those of ordinary GRBs. Thus the property of the extremely high $E_{\gamma,iso}$ would still be a unique property of Pop III VMS GRBs.

Given the propagation speed of the Poynting flux dominated jet inside the progenitor star, $\sim 0.2c$, deduced from MHD simulations \citep{barkov08,bromberg15}, the intrinsic duration of Pop III VMS GRBs $T/(1+z) \sim 10^4\;$s is sufficient to break through the star \citep[see also][]{meszaros01}. \cite{Suwa11} performed a semi-analytic calculation of the jet propagation inside the star with $R \sim 10^{13}\;$cm, which is an order of magnitude larger than our assumption, and showed that the jets can successfully break through the star \citep[see also][]{nagakura12}. While the jet is propagating inside the star, the energy of the jet is dissipated at the jet head and makes a thermally dominated cocoon between the jet and the stellar envelope \citep{begelman89,meszaros01,matzner03,bromberg11,bromberg15}. The cocoon can also be ejected after the jet breakout, and it can release thermal emission, which provide information about the progenitor star \citep{kashiyama13}. Indeed, the luminous supernova-like emission associated with Ultra-long GRB 111209A can be explained by the cocoon thermal emission \citep{nakauchi13} \citep[but see][]{greiner15}.

\subsection{Prompt emission}

We may roughly estimate the isotropic gamma-ray luminosity from equations (\ref{eq:Eiso}) and (\ref{eq:duration}) by $\sim E_{\gamma,iso}/[T/(1+z)]$ as
\begin{equation}
  L_{\gamma,iso} \sim 3 \times 10^{53}\;\epsilon_{\gamma} \eta_{-2} \alpha_{-1} \theta_{j,-1} R_{12}^{-\frac{3}{2}} \left(\frac{M_{\rm acc}}{300\;M_\odot}\right) \left(\frac{M_{\rm BH}}{300\;M_\odot}\right)^{\frac{1}{2}}\;{\rm erg}\;{\rm s}^{-1},
\end{equation}
where the notation $Q_x = Q/10^x$ in cgs units is adopted. This value is only one order of magnitude larger than that of ordinary GRBs.

We require the spectral model to examine the detectability of the prompt emission of Pop III VMS GRBs, although the prompt emission mechanism of GRBs is still under debate. Particularly for the magnetically dominated jets, it is very uncertain.

Even the magnetically dominated jets may have a subdominant thermal component, which may be released at the photosphere of the jets. The spectrum of this thermal emission can be estimated in a standard fireball model \citep{toma11}. For the photospheric thermal luminosity as $3 \times 10^{52}\;{\rm erg}\;{\rm s}^{-1}$ and $1+z = 20$, the spectral peak energy and the flux are estimated as $E_{p,obs} \simeq 70\;$keV and $F \simeq 7 \times 10^{-9}\;{\rm erg}\;{\rm cm}^{-2}\;{\rm s}^{-1}$. This emission is marginally detectable by the BAT detector on Swift.

An alternative way to estimate the spectral peak energy of the prompt emission is based on empirical laws claimed for observed ordinary GRBs \citep{Nakauchi12}. If the Poynting flux of the jets are efficiently converted to the radiation energy flux, the observed flux is $F \sim 7 \times 10^{-8}\;{\rm erg}\;{\rm cm}^{-2}\;{\rm s}^{-1}$ for $L_{\gamma,iso} \sim 3 \times 10^{53}\;{\rm erg}\;{\rm s}^{-1}$ and $1+z = 20$. Then if the spectral property obeys the empirical $E_p - L_{\gamma,iso}$ relation \citep{yonetoku10}, we have $E_{p,obs} \sim 80\;$keV, or if the spectral property obeys the empirical $E_p - E_{\gamma,iso}$ relation \citep{amati06}, we have $E_{p,obs} \sim 4\;$MeV. In both cases the prompt emission is detectable by currently operating satellites.

However, we have adopted $\eta \sim 10^{-2}$ for the above estimates, which correspond to the high end of the estimate of $\eta$ (see Section~\ref{subsec:GRBmodel}). For conservative values $\eta \sim 10^{-4} - 10^{-3}$, the radiation luminosity is orders of magnitude smaller, so that their detections by the current satellites are difficult. Next-generation satellites will be needed in such cases \citep{Suwa11,Nakauchi12}.

\subsection{Afterglow}

\begin{figure*}
  \begin{center}
  \includegraphics[width=0.8\textwidth]{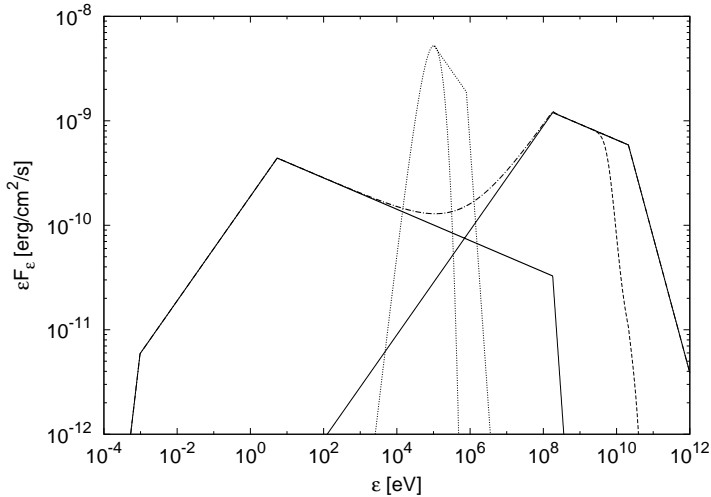}
  \caption{Example of the spectrum of a Pop III VMS GRB at the time when the jet activity ends. The physical parameters are $E_{iso} = 4 \times 10^{57}\;$erg, $T= 2.3\;$day, $1+z = 20$, $n = 1\;{\rm cm}^{-3}$, $\epsilon_B = 10^{-2}$, $\epsilon_e = 10^{-1}$, and $p = 2.3$. The dot-dashed line is the external shock spectrum, which consists of the synchrotron and SSC components (solid lines). The absorption due to extragalactic background light is expected to become significant above 7 GeV, as shown by the dashed line. The dotted line represents the prompt emission as a subdominant photospheric thermal component. See \cite{toma11} for more details.}
\label{fig:afterglow_spectrum}       
\end{center}
\end{figure*}

The external shock driven by the jet in the circumburst medium powers the afterglow, which can be studied independently of the prompt emission \citep{meszaros97,sari98}. The external shock amplifies the magnetic field in the shocked region via plasma and/or magnetohydrodynamic instabilities and accelerates the electrons in the shocked region to a power-law energy distribution. The accelerated electrons produce synchrotron and synchrotron-self-Compton (SSC) radiation as an afterglow. This afterglow model seems to be robust, since it can explain many of the late-time (i.e. since several hours after the burst trigger) multi-band afterglows so far, and simple extension of this model (e.g. continuous energy injection into the external shock) may explain many of the early-time afterglows \citep{liang07}.

Calculations of the external shock emission spectrum involve the parameters $E_{iso}$ (i.e. the total jet energy minus $E_{\gamma,iso}$), the number density $n$ of the circumburst medium, the fractions $\epsilon_B$ and $\epsilon_e$ of the thermal energy in the shocked region that are carried by the magnetic field and the accelerated electrons, respectively, and the index $p$ of the energy spectrum of the accelerated electrons. The latter three microphysical parameters have been constrained by model-fitting the late-time afterglows as $10^{-5} \la \epsilon_B \la 10^{-1}$, $\epsilon_e \sim 10^{-1}$, and $p \sim 2.3$ \citep[e.g.][]{panaitescu02,wijers99}.

\begin{figure*}
  \begin{center}
  \includegraphics[width=0.8\textwidth]{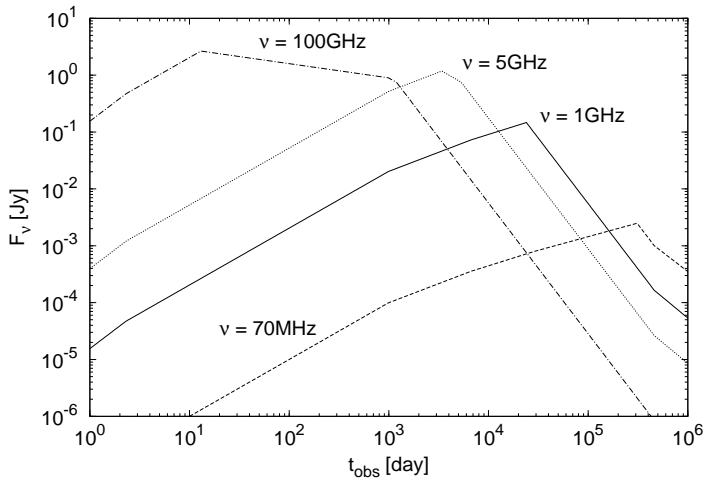}
  \caption{Radio light curves at frequencies, 100 GHz, 5 GHz, 1 GHz, 70 MHz, of a Pop III VMS GRB at $1+z = 20$ with the same physical parameters as Figure~\ref{fig:afterglow_spectrum}. See \cite{toma11} for more details.}
\label{fig:afterglow_radio}       
\end{center}
  \end{figure*}

Figure~\ref{fig:afterglow_spectrum} is an afterglow spectrum at the time $t \sim T$ from the burst trigger (i.e. the early brightest phase of the afterglow) calculated for Pop III VMS GRBs by \cite{toma11}. The parameters are $E_{iso} = 4 \times 10^{57}\;$erg, $T= 2.3\;$day, $1+z = 20$, $n = 1\;{\rm cm}^{-3}$, $\epsilon_B = 10^{-2}$, $\epsilon_e = 10^{-1}$, and $p = 2.3$. We found that the flux at the IR frequency is sufficient to estimate the redshift. (\cite{macpherson13} discuss the afterglow detectability with JWST and SPICA more systematically for broad range of parameters.) The Swift XRT and Fermi LAT can detect the X-ray and gamma-ray radiation of this afterglow, which may constrain the total energy scale $E_{iso}$ and the number density $n$ of the circumburst medium. The constraint on $n$ would provide invaluable information about the environment and the radiation feedback of Pop III VMSs at the phase prior to their explosions \citep{kitayama04,whalen04,Wang12}. A caveat is that in order to perform X-ray observation, one of the wide-field detectors, e.g. Swift BAT and Fermi LAT, has to be triggered by the prompt emission or the afterglow itself.

\citet{toma11} also found that the afterglows in the radio band are so bright that they might be detected by survey observations with current radio telescopes. Figure~\ref{fig:afterglow_radio} is the light curves at various radio frequencies of the afterglow with the same physical parameters as Figure~\ref{fig:afterglow_spectrum}. This shows that the radio afterglows of Pop III VMS GRBs can be very bright with very long durations. These may be point sources with very long variability times in the radio sky. Deep radio surveys might detect them or can constrain the rate of the Pop III VMS GRBs \citep{toma11,desouza11,ghirlanda13,macpherson15}. Furthermore, such bright radio sources could be useful for 21 cm absorption line searches \citep{furlanetto02,ioka05,toma11,ciardi15}, although we should note that the brightness of the afterglow highly depends on the uncertain parameters discussed in Section~\ref{subsec:GRBmodel}, similarly to the prompt emission.

The external shock can also accelerate protons to the non-thermal energy distribution, which can produce high-energy neutrinos via $p\gamma$ interaction. The flux of these neutrinos can be high in the $10-100$ PeV energy range \citep{gao11} \citep[see also][]{berezinsky12}.

\section{Outlook}

What then are the prospects for probing the end of the cosmic dark ages with GRBs, and what are key
challenges ahead? For the time being, the {\it Swift} 
satellite, with its on-board X-ray and optical telescopes for the rapid localization of gamma-ray transients in the sky, will remain a good facility for catching high-redshift GRBs with $z \ga 6-9$. The next mission, SVOM, with similar capabilities as {\it Swift}, is now being prepared to follow it. Several other satellite missions targeting higher-redshift GRBs, such as JANUS and Lobster, have also been proposed. Increasing the statistics of high-redshift GRBs, and detecting peculiar sources suggestive of VMS progenitors, will constrain theories of the formation and evolution of Pop~III stars, and their GRB production mechanism, as discussed in this article. Wide-field, space-based gamma-ray/X-ray detectors are required to keep operating not only for high-redshift observations but also for upcoming gravitational wave astronomy, which requires electromagnetic counterparts for robust confirmation. The deep surveys in the radio wavelengths may also be helpful for Pop~III GRB searches.

From the theory side, we cannot yet make any robust predictions for the final masses (i.e., the IMF), radii, rotation speeds, magnetic field strengths and its configurations, and the binarity of Pop~III stars either at the time of formation or at the final evolution stage before the explosion. Similarly, during the GRB explosion phase, the mechanisms of driving relativistic jets and releasing high-energy emissions are still subject to considerable uncertainty. However, numerical simulations have been steadily extending our understanding of some of those ingredients, and they will keep playing a crucial role in theoretical study.

It is clearly important to improve our understanding of the Pop~III GRB formation channel, in particular regarding the rotation state of the progenitor stars, and the characteristics of tight binaries. The challenge here is two-fold. We need simulations with even higher resolution, to push into the regime of possible binary overflow phenomena (Roche-lobe and common envelope), while maintaining the large-scale cosmological boundary conditions. The second challenge is improved physics, in particular inclusion of magnetohydrodynamic effects. The latter may crucially impact both the fragmentation properties of the primordial gas, and the rotation rates of the resulting Pop~III stars.

The effects of rotation, magnetic fields and binary interactions are also critical in the evolution of Pop III GRB progenitor stars. Furthermore, the rapid rotation and strong magnetic field are key ingredients for producing relativistic jets via the BZ process. Our understanding of these physical processes in massive star evolutions has been greatly improved during the last decade both observationally and theoretically, but uncertainties regarding the transport processes of angular momentum and chemical species inside stars still remain large. In particular, we still lack direct observations of very metal poor massive stars, which makes it difficult to test stellar evolution theory of massive Pop III stars. We hope to overcome this difficulty with systematic studies on massive star populations in very metal poor environments like in I Zwicky 18 \citep[e.g.][]{szecsi15}.

Another key uncertainty relates to theoretical predictions for the rate of Pop~III bursts.
Even if we can narrow down the uncertainty in the Pop~III GRB formation efficiency, as outlined in the previous paragraph, we still have to contend with the poorly
constrained Pop~III star formation rate density. This rate is only known within 
2 to 3 orders of magnitude, mostly because of our incomplete understanding
of the (radiative and SN) feedback effects that regulate the primordial star formation process.
Again, improved cosmological simulations with ever more realistic feedback
implementations promise to provide more robust determinations in the next few
years. To give a rough ballpark impression of current, state-of-the-art
predictions for the Pop~III GRB rate at $z\gtrsim 6$: a robust upper
limit should be $\sim 10$\,yr$^{-1}$\,sr$^{-1}$ \citep{salvaterra11}, and a
more realistic limit may be
$0.1$\,yr$^{-1}$\,sr$^{-1}$ \citep{campisi11}. Again, the hope is that progress
in our understanding of both the Pop~III GRB progenitor physics and the 
global star formation rate density is achievable in the near future.


%
%

\begin{acknowledgements}
  K.T. thank P. M\'{e}sz\'{a}ros, K. Ioka, Y. Suwa, and D. Nakauchi for useful
comments. This work is partly supported by JSPS Grants-in-Aid for Scientific
Research 15H05437 (K.T.), JST grant ``Building of Consortia for the Development
of Human Resources in Science and Technology'' (K.T.) and NSF grant AST-1211729
(V.B.).  SCY was supported  by Basic Science Research (2013R1A1A2061842)
program through the National Research Foundation of Korea (NRF)
\end{acknowledgements}

\bibliographystyle{aps-nameyear}      
\bibliography{example}                
\nocite{*}


\end{document}